\documentclass[useAMS,usenatbib, twocolumn]{mn2e}
\usepackage{graphicx}
\usepackage{lscape}
\usepackage{longtable}
\usepackage{rotating}
\usepackage{colordvi}
\usepackage{color}
\usepackage[english]{babel}
\usepackage{amsmath}
\usepackage{amssymb}
\usepackage{amsfonts}
\usepackage{epsfig}
\usepackage{subfig}
\usepackage{textcomp}
\usepackage{natbib}
\usepackage{keyval}
\def\1{\'\i}
\def\gsim{\mathrel{\vcenter{\hbox{$>$}\nointerlineskip\hbox{$\sim$}}}}

\title[SZ/X-ray scaling relations 
using X-ray data and Planck Nominal maps.]{SZ/X-ray scaling relations 
using  X-ray data and Planck Nominal maps.}
\author[De Martino \& Atrio-Barandela]{
I. De Martino$^{1}$, F. Atrio-Barandela$^{2}$ \\ 
$^{1}$ Department of Theoretical Physics and History of Science, 
University of the Basque Country UPV/EHU, Faculty of Science\\
and Technology, Barrio Sarriena s/n, 48940 Leioa, Spain;\\ 
$^{2}$ F\1sica Te\'orica, Universidad de Salamanca, 37008 Salamanca, Spain;\\
email: ivan.demartino1983@gmail.com; atrio@usal.es
}
\begin{document}
\date{Accepted xxx. Received yyy; in original form zzz}
\pagerange{\pageref{firstpage}--\pageref{lastpage}} 
\pubyear{2015}
\maketitle
\label{firstpage}

\begin{abstract}
We determine the relation between the
Comptonization parameter predicted using X-ray data $Y_{C,Xray}$ and the
X-ray luminosity $L_X$, both magnitudes derived from
ROSAT data, with the Comptonization parameter $Y_{C,SZ}$
measured on {\it Planck} 2013 foreground cleaned Nominal maps.
The 560 clusters of our sample includes clusters with masses 
$M\ge 10^{13}M_\odot$, one order of magnitude smaller 
than those used by the Planck Collaboration in a 
similar analysis. It also contains eight times more 
clusters in the redshift interval $z\le 0.3$. 
The prediction of the $\beta=2/3$ model convolved with
the Planck antenna beam agrees with the anisotropies
measured in foreground cleaned Planck Nominal maps within the X-ray emitting
region, confirming the results of an earlier analysis \citep{atrio2008}. 
The universal pressure profile overestimates the signal by a  15-21\%
depending on the angular aperture. We show that the discrepancy is
not due to the presence of {\it cool-core} systems but it is an
indication of a brake in the $L_X-M$ relation towards low mass
systems. We show that relation of the Comptonization parameter
averaged over the region that emits 99\% of the X-ray flux and 
and the X-ray luminosity is consistent with the predictions of the 
self-similar model. We confirm previous findings that the scaling 
relations studied here do not evolve with redshift within the range 
probed by our catalog.
\end{abstract}

\begin{keywords}
galaxies: clusters: general - X-rays: galaxies: clusters - cosmology: observations
\end{keywords}

\section{Introduction.}

Clusters of galaxies are the largest virialized structures in the Universe
first observed as concentrations of optical galaxies.
Compression and shock-heating processes raise 
the temperature of the Intra-Cluster Medium (ICM) to $T_X\sim 1-10$keV
and clusters can be observed through their X-ray emission and their 
thermal Sunyaev-Zeldovich (TSZ, \citet{tsz}) distortion of the Cosmic Microwave 
Background (CMB).  The {\it self-similar} model predicts simple scaling relations 
between cluster observables and their mass \citep{kaiser_86}. More specifically, 
hydrodynamical N-body simulations have shown that the TSZ signal integrated over 
the cluster volume scales with the cluster mass \citep{white2002, daSilva2004, 
motl2005, nagai2006, wik2008, Aghanim2009}. X-ray properties are also 
related to the cluster mass and gas temperature \citep{Melin2011}. 
Scaling relations have been determined observationally but 
their form does not necessarily coincide with the prediction of the self-similar
model \citep{Voit2005, arnaud2005, arnaud2007, pratt2009, Vikhlinin2009}.
Therefore, non-gravitational processes such as mergers, cooling and energy 
injection from Active Galactic Nuclei (AGN) have a significant contribution
to the equilibrium state of clusters and scaling relations can be used 
to test the physics of clusters of galaxies \citep{Bonamente2008, 
Marrone2009, arnaud2010, Melin2011, Andersson2011, Comis2011}.
The redshift evolution of the scaling relations seems to follow 
the self-similar prediction \citep{bower1997,Maughan2006}, suggesting that
cluster properties evolve with the density of the Universe. 
Large-scale surveys of the TSZ effect such as those carried out by the 
Atacama Cosmology Telescope (ACT, \citet{Kosowsky2003}) the South Pole 
Telescope (SPT, \citet{carlstrom2011}) and the {\it Planck} satellite 
\citep{planck_01} have helped to constrain cosmological parameters using 
SZ clusters \citep{Vanderlinde2010,Sehgal2011,planck13_XX,planck15_XXIV}
and to establish scaling relations between X-ray magnitudes and TSZ 
measurements \citep{pratt2006, Maughan2007, pratt2009, 
arnaud2010, Comis2011, Andersson2011, komatsu2011, Rozo2012, bolocam_15}. 
The effectiveness of clusters as cosmological probes depends
on obtaining reliable mass estimates. In this respect, the scatter on
the scaling relations needs to be understood in order to account for 
observational biases. The scatter includes statistical errors, systematic
biases and the intrinsic differences between clusters. The intrinsic scatter 
is dominated by the cluster cores \citep{oHara2006,Chen2007} since 
\emph{cool-core} (CC) clusters deviate from self-similarity of the 
\emph{no cool-core} (NCC) population.  
The difference in the cluster population have been thought to be responsible
for discrepancies between numerical predictions and observations.
\citet{komatsu2011} found that the universal pressure profile 
\citep{arnaud2010} overestimated the TSZ amplitude by a 30\%. 
Since the universal pressure profile agreed with TSZ data for CC 
but disagreed for the NCC clusters, it was thought that the
excess was related to the dynamical state of the ICM. Nevertheless, 
subsequent analysis did not find any evidence of this discrepancy 
\citep{Melin2011, Andersson2011, Rozo2012}. This was further confirmed by
the Planck Collaboration using a sample of $\sim 1600$ clusters
\citep{planck_early_X,planck_early_XI}. In this study, the TSZ anisotropy 
was individually measured for each object but, in order to maximize 
the statistical significance of the result, the signal was averaged 
in bins of X-ray luminosity. The Planck Collaboration 
also carried out an analysis of scaling relations using a sample of 62 clusters
(later extended to 78) without binning their properties 
\citep{planck_early_XI,planck13_XXIX}. In this paper we extent this later analyses 
using a sample of 560 clusters with redshifts $z\le 0.3$ that includes systems with
masses one order of magnitude smaller than the latter Planck sample. Briefly, 
in Sec.~2 we review cluster pressure profiles models,
in Sec.~3 we describe our cluster catalog and the CMB data used in this study and
in Sec.~4 we discuss the scaling relations to be determined, the regression
routines used and the associated errors. Finally, in Sec.~5 we present our 
results and in Sec.~6 we summarize our conclusions.
\begin{table*}
\begin{center}
\begin{tabular}{|l|ccccc|}
\hline
 {\bf Reference} & $P_0$ & $c_{500}$ & $\gamma$ & $\alpha$ & $\beta$ \\
 \hline
 \citet{arnaud2010} & 4.291$h^{-3/2}$ & 1.177 & 0.3081 & 1.0510 & 5.4905\\
 \citet{planck_int_05} &  6.41 & 1.81 & 0.31 & 1.33 & 4.13\\
 \hline
\end{tabular}
\caption{The universal pressure profile parameters determined
by \citet{arnaud2010} and \citet{planck_int_05}.}\label{tab:gnfw_pars}
\end{center}
\end{table*}

\section{The Integrated Comptonization Parameter.}\label{sec:profile}

The TSZ effect is a distortion of the CMB black-body spectrum produced when 
CMB photons are scattered off by the free electrons of the ICM.
The TSZ effect is independent 
of redshift and it is usually expressed in terms of the Comptonization parameter 
$y_c=(k_B\sigma_T/m_ec^2)\int_l T_Xn_e dl$, where $n_e, T_X$ 
are the electron density and temperature along the line of sight $l$, 
$\sigma_T$ the Thomson cross section, $m_e$ the electron rest mass and
$c$ the speed of light.
The temperature anisotropy $\Delta T_{TSZ}(\hat{n})= T_0G(\nu)y_c$, with
$T_0$ is the CMB black-body temperature and $G(\nu)$ is the 
spectral dependence of the TSZ effect, that is different from that of 
any other known astrophysical foreground. In the non-relativistic limit 
$G(\nu)= \tilde{\nu}{\rm coth}(\tilde{\nu}/2)-4$, where 
$\tilde{\nu}=h\nu/k_BT_X$ is the reduced frequency and $h,\; k_B$ are the Planck
and Boltzmann constants, respectively.
In this work we will include relativistic corrections up to fourth order 
in the electron temperature (\citet{itoh1998, nozawa1998, nozawa2006}).
The SZ effect integrated over the solid angle subtended by the cluster is
\begin{align}
&Y_C=\int y_cd\Omega=D_A^{-2}(z)\int y_cdA\nonumber\\[0.3cm]
& =\frac{k_B\sigma_{\rm T}}{mc^2}D_A^{-2}(z)
\int_VT_{\rm e}n_{\rm e}dV=\frac{\sigma_{\rm T}}{mc^2}D_A^{-2}(z)\int_VP_{\rm e}dV,
\label{eq:Y}
\end{align}
where $D_A(z)$ is the angular diameter distance, $dA$ the projected area and
$dV=dldA$ the volume element. The integrated
Comptonization parameter is dimensionless;
i.e., it is measured in units of solid angle, generally in $(arcmin)^2$. An
alternative convention is to use $D^2_A~Y_C$ and express this magnitude in units 
of Mpc$^2$. 

The pressure profile of the hot ICM was initially des\-cribed by an
spherically symmetric isothermal gas with the electron density profile given by
the $\beta$ model \citep{cavaliere1976}
\begin{equation}\label{eq:beta-model}
n_e(r)=n_{e,0}\left[1+\left(\frac{r}{r_c}\right)^2 \right]^{-3\beta/2}, 
\end{equation}
where the cluster core radius $r_c$, central electron density $n_{e,0}$ 
and slope $\beta$ are determined from observations. By fitting 
the cluster X-ray surface brightness \citet{jones1984} estimated 
$\beta=0.6-0.8$. The TSZ emission predicted for a cluster sample 
using the fiducial value $\beta=2/3$, convolved with the WMAP 
antenna beam has been shown by \citet{atrio2008} to agree with
the measured anisotropy in WMAP 3yr maps within the region emitting 
99\% of the X-ray flux. 

Based on the Navarro-Frenk-White (NFW) 
dark matter profile \citep{nfw1997} and using the the results of N-body simulations, 
\citet{nagai2007} proposed the {\it dimensionless} pressure profile
\begin{equation}\label{eq:prof_nagai}
p(x)=\frac{P_0}{(c_{500}x)^\gamma[1+(c_{500}x)^\alpha]^{(\beta-\gamma)/\alpha}},
\end{equation}
where $x=r/r_{500}$ is the distance from the cluster center in units of 
the radius at which the mean overdensity of the cluster is 500 times the 
critical density, $c_{500}$ is the gas concentration parameter at $r_{500}$,
$(\gamma,\alpha,\beta)$ are the central, intermediate and outer slopes and
$P_0$ is given in Table~\ref{tab:gnfw_pars}.
\citet{arnaud2010} assumed the profile of eq.~(\ref{eq:prof_nagai}) to be universal 
and used X-ray data from a sample of 33 clusters at redshift $z<0.2$ to constrain 
the model parameters. The {\it dimensional} electron pressure profile derived 
from that data was
\begin{align}
P_e(x)&=3.36~h^2E^{8/3}(z)\times\nonumber\\[0.3cm]
&\times\left[\frac{M_{500}}{2.1\times 
10^{14}h^{-1}M_\odot}\right]^{2/3+\alpha_p+\alpha'_p} p(x)~[{\rm eV~cm^{-3}}],
\label{eq:universal-profile}
\end{align}
with $\alpha_p=0.12$, and $\alpha'_p(x)=0.1−(\alpha_p+0.10)
(x/0.5)^3(1.0+(x/0.5)^3)^{-1}$. 
In this expression $h=H_0/[100{\rm km s^{-1}Mpc^{-1}}]$ is the reduced
Hubble constant with $H_0$ its current value. 

The parameters 
of eq.~(\ref{eq:prof_nagai}) have also been determined
using the TSZ anisotropy of 62 nearby clusters,
measured in the 2013 {\it Planck} CMB data \citep{planck_int_05}. 
To test how well the model predictions agree with the measured anisotropy,
in this article we will analyze if the amplitude of the TSZ
effect predicted from the X-ray data convolved with the beam at each
{\it Planck} frequency agrees with the measured anisotropy using
the isothermal $\beta$ model of eq.~(\ref{eq:beta-model}) and the pressure profile 
of eq.~(\ref{eq:universal-profile}) with the two sets of parameters given 
in Table~\ref{tab:gnfw_pars}.

\section{Data.}

We use a sample of 560 X-ray clusters and the {\it Planck} Nominal data released in 
2013\footnote{Data downloaded from {\it http://www.cosmos.esa.int/web/planck}}
to determine the X-ray/SZ scaling relation. Our work differs from previous
studies in that all magnitudes used in our analyses have been derived from 
observations, except the X-ray temperature and $r_{500}$ scale that 
were themselves obtained from scaling relations.

\subsection{X-ray Cluster Catalog.}\label{sec:xray_catalog}

The cluster catalog was compiled from three ROSAT X-ray flux limited surveys: the 
ROSAT-ESO Flux Limited X-ray catalog (REFLEX, \citet{bohringer04}), 
the extended Brightest Cluster Sample (eBCS, \citet{ebeling98, ebeling00})
and the Clusters in the Zone of Avoidance (CIZA, \citet{ebeling02}).
These three samples differ in selection techniques, flux measuring
algorithms and are affected by different systematic effects. To construct
a homogeneous all-sky sample, the different selection technique and 
the flux determination method employed have to be taken into account
to guarantee that all three samples are complete to the same depth.
A full discussion of the method used to combine the individual catalogs into
a homogeneous all-sky sample is given in \cite{kocevski-ebeling06} and
is briefly summarized here. First, the flux is recomputed using ROSAT
All Sky Survey (RASS) data.
The centroid of the cluster X-ray emission is determined and 
the X-ray count rate is computed taking into account the local RASS 
exposure time.
The X-ray background is determined from an annulus of radius
1 and 1.5$h^{-1}$Mpc around the cluster centroid and subtracted from the 
measured counts. The resulting count rates are 
deconvolved from the telescope Point Spread Function (PSF) and
converted to unabsorbed fluxes in the [0.1-2.4]keV band. For the RASS, 
the PSF is the weighted average of the PSF's at all off-axis angles 
\citep{ebeling98}.  Clusters whose emission is dominated by a point source
were removed and a cut of $F_x[0.1-2.4{\rm keV}]\ge 3\times 10^{-12}$ 
ergs cm$^{-2}$ s$^{-1}$ was applied.  The merged catalog
contains 782 clusters with well measured positions, spectroscopic redshifts,
X-ray fluxes in the [0.1-2.4]keV band and angular extents of the region 
emitting 99\% of the X-ray flux, hereafter $\theta_X$. Of those,
only 623 clusters survive the point-source and the Planck galactic masks.
Foreground contamination reduced the total number of clusters used in
this study to $N_{cl}=560$.

All the clusters in our sample were fitted to an isothermal $\beta$ model
\citep{cavaliere1976}. If $S(r)$ is the projected surface brightness
distribution, then $S(r) = S_0 \left[
1+(r/r_c)^2\right]^{-3\beta+1/2}$ where $S_0$, $r_c$, and $\beta$ are
the central surface brightness, the core radius, and the $\beta$
parameter characterizing the profile, respectively.
Due to the low angular resolution of the RASS the surface
brightness of our clusters is poorly sampled except for 
nearby clusters. The correlation between $r_c$ and $\beta$
introduces further uncertainties and makes the results for
both parameters sensitive to the radius of the cluster
chosen to fit the model. Due to these limitations, we
take $\beta=2/3$, the canonical value \citep{jones1984}. 
Reassuringly, the values of $r_c$ derived from the data
agree with the values derived from the $L_X-r_c$ relation
determined by \citet{reiprich_99}.  
Cluster luminosities and electron temperatures are
used to determine central electron densities.
The ICM temperature is derived from the bias corrected
$L_{\rm X}[0.1-2.4keV]-T_{\rm X}$ relation of \cite{lovisari_15}.

From the RASS data, X-ray luminosities are measured within a radius of angular size 
$\theta_X$, and are k-corrected to rest frame $[0.1-2.4]$ keV from the 
REFLEX/CIZA/eBCS surveys. Conversions between angular extents and 
physical dimensions are made using the $\Lambda$CDM model 
with {\it Planck} measured parameters \citep{planck13_XVI}. 
Errors are due to Poisson noise in the
number of the photons detected for each cluster and are, at most,
20\%\footnote{H. Ebeling, private communication}.  By fitting a
$\beta=2/3$ model to the X-ray surface density, the core radii ($r_c$) and 
central electron density ($n_{e,0}$) have been determined. 
To compare with 
previous analyses, we also evaluate our scaling relations at $r_{500}$. We derive
this scale from the $r_{500}-L_X$ relation of \citet{bohringer07}.
From the latter magnitude we define the angular size $\theta_{500}=r_{500}/D_A(z)$ 
and mass scale $M_{500}=({4\pi}/{3})500 \rho_c(z)r_{500}^3$, where 
$\rho_c(z)$ is the critical density at the cluster redshift. 
These clusters are located within the redshift interval $z=[0,0.3]$, have luminosities 
$L_X=[0.3 - 22.5]\times 10^{44}$ erg/s that correspond to $T_X=[0.87 - 11.5]$ keV, 
and $M_{500}=[0.2-14.7]\times 10^{14}$M$_\odot$. 
By comparison, the Planck Collaboration study used 78 individual clusters with
$z=[0.0, 0.5]$, $T_X=[3-14]$keV and $M_{500}=[2-20]\times 10^{14}M_\odot$. 
The larger number of clusters and the wider mass range will allow us
to test the accuracy with which the cluster pressure profiles of
eqs.~(\ref{eq:beta-model}) and (\ref{eq:universal-profile}) fit the data and how 
well the average properties of the cluster population are described
by the self-similar model.

\subsection{Foreground cleaned Planck Nominal Maps.}\label{sec:cleaning}

{\it Planck} data were originally released in 2013 in a Healpix format 
with resolution $N_{side} = 2048$ \citep{gorski2005}. The Nominal maps 
contained foreground emissions from galactic dust, CO lines, synchrotron, 
point sources and extended infrared sources.  The angular resolution 
of the Low Frequency Instrument channels is $\theta_{FWHM}>13'$, larger
than the angular extent of the clusters in our sample, so we will
restrict our analysis to the High Frequency Instrument data.
Prior to compute the TSZ anisotropies we clean the data from 
foreground and cosmological contributions as described in \citet{demartino2015}.
First, we subtract the intrinsic CMB and kinematic SZ anisotropies using 
the LGMCA CMB template \citep{bobin2013, bobin2014}.
Thermal dust emission is subtracted using the 857~GHz channel as a
dust-template \citep{planck13_XI, planck13_XII}.  We clean this
contribution on sky patches $\mathcal{P}(\nu, i)$ centered on each 
cluster $i$ at each frequency $\nu$ following \cite{diego2002}. 
The CO emission at 100 and 217~GHz is removed using the 
Type 2 maps described in \cite{planck13_XIII}.  We 
used the PCCS-SZ-Union mask to excise point sources and mask the 
residual Galactic Plane emission \citep{planck13_XXVIII, planck13_XXIX}. 
Temperature fluctuations ($\delta \bar{T}$) are measured by averaging the 
anisotropy over a disc of size $\theta$ on the foreground cleaned patches.
The method is not fully effective and 
we rejected those clusters for which $Y_{C,500}<0$, corresponding to high 
redshift and low luminosity systems with high levels of foreground residuals.
In total, $N_{cl}=560$ clusters were used in this study.

In addition to the TSZ signal of interest, $Y_C(x)G(\nu)$, these foreground 
cleaned patches $\mathcal{P}$ contain instrumental noise $N(\nu,x)$ and 
some degree of CMB and 
foreground residuals.  To estimate error bars, we placed discs the size of
each cluster on 1,000 random positions.
The patch covered by each random disc is cleaned using the
procedure described above and then the mean temperature anisotropy on
the disc is computed. The process is repeated for all clusters.
To avoid overlapping these random clusters with the real population, 
we mask an area of one degree radius around all known clusters.
The error bar associated with the measured anisotropy of a cluster is  
\begin{equation}
\label{eq:SP_error}
\sigma^2(\theta,\nu_i)= 
\langle[\delta\bar{T}(\theta,\nu_i)-\mu(\theta,\nu_i)]^2\rangle^{1/2},
\end{equation}
where $\mu(\theta,\nu_i)=\langle\delta \bar{T}(\theta,\nu_i)\rangle$
and averages are taken over the 1,000 random positions. 
Our cleaning method performs better for small apertures. The average relative
error was $\sim 13\%$ at $\theta_{500}$ and it grows to $\sim 40\%$ at
$2\theta_{500}$. Therefore, we will perform our analysis on apertures
equal or smaller than $\theta_{500}$.

The integrated Comptonization parameter $Y_C(\theta)$ is measured using 
two different methods:\\

\parindent=0cm (A) At a cluster location and for all frequencies we define 
the signal as $Y_{C}(\theta,\nu)=\delta \bar{T}/(T_0 G(\nu))$ and the 
associated error as $\sigma_{Y_C}(\theta,\nu)=\sigma(\theta,\nu)/(T_0G(\nu))$.
Since $G(217{\rm GHz})\approx 0$, $Y_{C}(\theta,{\rm 217 GHz})$ will be dominated
by the errors (see below), the Comptonization parameter will be computed as the 
weighted average over all frequencies except 217~GHz. It is given by
\begin{align}
&\bar{Y}_{C,\nu}(\theta)=\sigma_{\bar{Y}_{C,\nu}}^2(\theta)\Sigma_{\nu} 
\left[\frac{Y_C(\theta,\nu)}{\sigma^2_{Y_C}(\theta,\nu)}\right]; \nonumber\\
&\sigma_{\bar{Y}_{C,\nu}}^{-2}(\theta)=\Sigma_{\nu}\sigma^{-2}_{Y_C}(\theta,\nu).
\label{eq:weighted_average}
\end{align}
This expression does not include the negligible error on the CMB blackbody 
temperature $T_0$.\\

\parindent=0cm (B) The foreground cleaned patches at all frequencies
are combined using the Internal Linear Combination (ILC) method described in
\citet{delabrouille2009} and the Comptonization parameter is measured directly
on the combined map.  
To illustrate how the ILC technique is applied to estimate the TSZ emission,
we assume that the residual cosmological signal $B(\nu)T(x)$ dominates
over the foreground residuals. Then, the anisotropy in each patch will
have the following components
\begin{equation}
\mathcal{P}(\nu,x)=Y_C(x)G(\nu)+B(\nu)T(x)+N(\nu,x).
\end{equation}
Following \citet{remazeilles2011}, we can estimate the TSZ emission 
in each patch as $\hat{Y}_C(x)=w(\nu_i)\mathcal{P}(\nu_i, x)$. The weights
$w(\nu)$ are obtained by minimizing 
$\chi^2=N_{\rm pix}^{-1}\sum_x\left(\hat{Y}_C(x)-\langle\hat{Y}_C\rangle\right)^2$ 
where $N_{\rm pix}$ is the number of pixels of each patch. The weights 
satisfy $\Sigma w(\nu_i)G(\nu_i)=1$, $\Sigma w(\nu_i)B(\nu_i)=0$ and are
given by
\begin{equation}
w(\nu_i)=\frac{\left(B_k\hat{R}^{-1}_{kl}B_l\right)G_j\hat{R}^{-1}_{ij}-
\left(G_k\hat{R}^{-1}_{kl}B_l\right)B_j\hat{R}^{-1}_{ij}}
{\left(G_k\hat{R}^{-1}_{kl}G_l\right)\left(B_m\hat{R}^{-1}_{mn}B_n\right)- 
\left(G_k \hat{R}^{-1}_{kl} B_l \right)^2} .
\end{equation}
Here $\hat{R}_{ij}=N_{\rm pix}^{-1}\sum_p\left(T_i(p)-\langle T_i\rangle\right) 
\left( T_j(p)-\langle T_j \rangle \right)$ is the empirical covariance matrix
computed on the foreground cleaned random patches and
the indices $i,j,k$ run over the frequencies $\nu=[100,143,217,353]$GHz.
Like before, the Comptonization parameter $\bar{Y}_{C,ILC,\theta}$ is obtained by
averaging on a disc of the radius $\theta$; the associated error 
is estimated using the ILC weights from 1,000 foreground cleaned patches
placed randomly outside the known cluster positions.

\begin{figure*}
\epsfxsize=1\textwidth \epsfbox{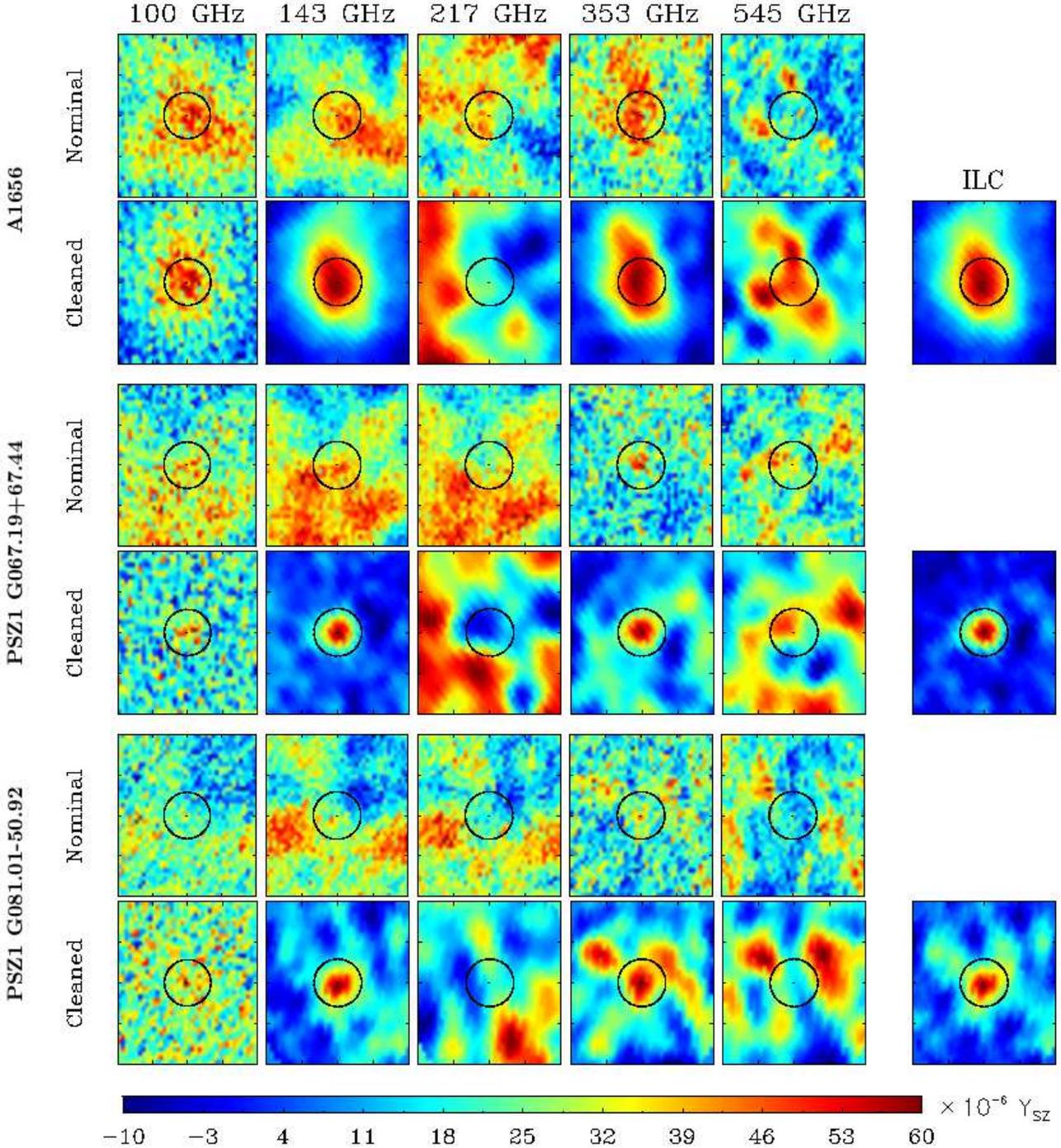}
\caption{Planck Nominal and  foreground cleaned patches
centered on the position of A1656, PSZ1 G067.19+67.44, and PSZ1 G081.01-50.92.
at 100-545~GHz, and  in the ILC foreground cleaned patch. 
The angular size of the patches is 1\textdegree$\times$1\textdegree. 
In each clusters, the black circle corresponds to a disc of radius
$\theta_{500}$.}\label{fig1}
\end{figure*}

To demonstrate how efficiently our two pipelines remove foregrounds and the 
different effect on the low redshift and extended, the intermediate
and the high redshift and compact clusters, in Fig.~\ref{fig1} we show the 
original Planck Nominal maps and the foreground cleaned patches in 
units of $T_0G(\nu)$ for each frequency. In these units, the TSZ anisotropy 
does not change sign but the CMB residuals do. 
Patches subtend a solid angle of 1\textdegree$\times$1\textdegree.
We selected three {\it Planck} clusters: A1656 (Coma) with redshift 
$z = 0.023$ and an angular extent of $\theta_{500} = 48.1^\prime$,
PSZ1 G067.19+67.44 with $z = 0.1712$ and $\theta_{500} = 9.34^\prime$
and  PSZ1 G081.01-50.92 $z = 0.2998$ and $\theta_{500} = 5.45^\prime$. 
For each cluster, we also present the ILC reconstruction of the TSZ signal
of the same data.  While the TSZ emission is zero at 217~GHz, it 
dominates over the foregrounds residuals at all other frequencies
except at 545 GHz, where dust residuals are still the dominant contribution.
Therefore, this channel will not be used to avoid biasing the results.

\begin{figure*}
\epsfxsize=1\textwidth \epsfbox{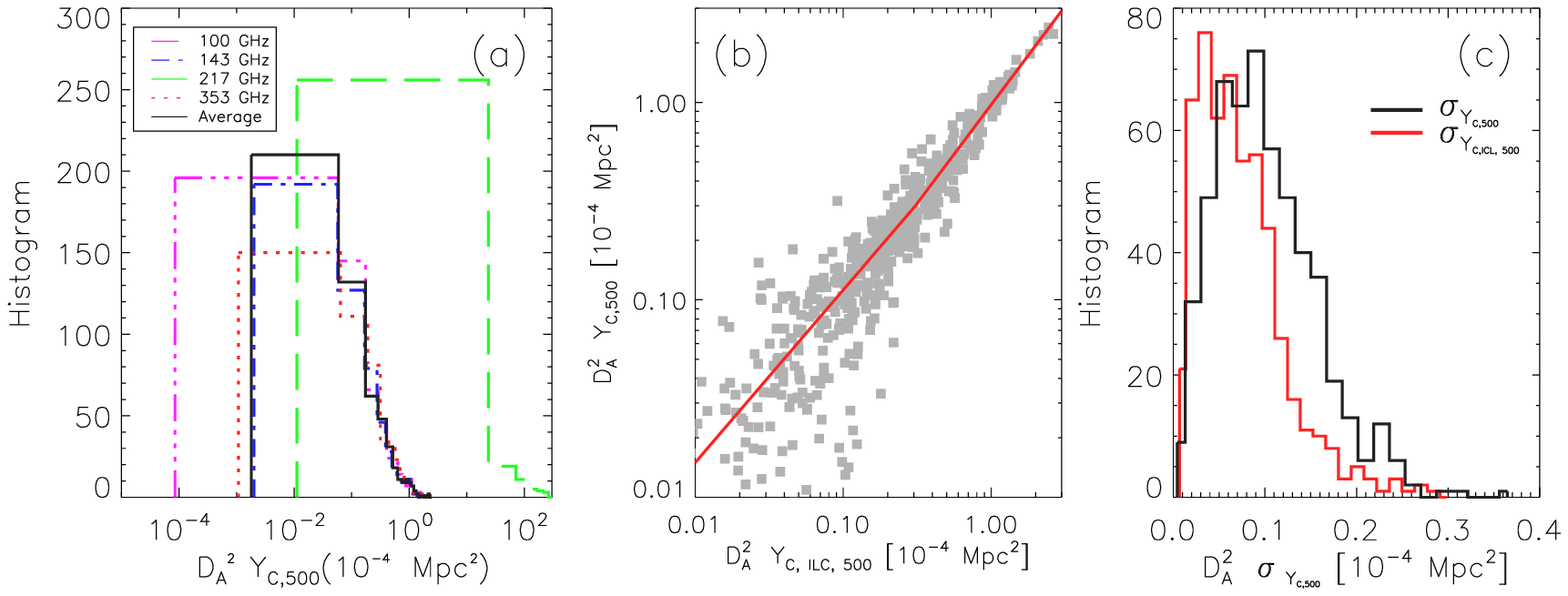}
\vspace*{-6.0cm}
\caption{(a) Histograms of the Comptonization parameters measured on the 
cleaned patches centered at the cluster positions. $\bar{Y}_{500}$ was computed on 
discs of size $\theta_{500}$ from 100 to 353~GHz using
eq.~\eqref{eq:weighted_average}; (b) Comparison of the Comptonization parameter
measured in the ILC map and measured by combining frequencies; the red solid line 
represents the best linear fit: $\bar{Y}_{\nu, 500}=A+B\bar{Y}_{ILC,500}$, 
with parameters $A=0.001\pm0.007$ and $B=0.97\pm0.03$. 
(c) Histogram of the errors on the measured Comptonization parameters
using the ILC (red line) and the combined frequencies method (black line).}\label{fig2}
\end{figure*}

Fig.~\ref{fig2} illustrates that the estimated Comptonization parameter
is independent of our foreground cleaning technique and estimation
method. In Fig.~\ref{fig2}a, we plot the distribution of the 
Comptonization parameter measured on discs of angular radius
$\theta_{500}$ at difference frequencies, $Y_{C,500}$, and its average
over all channels. All measured values are very similar except those 
at 217~GHz (long-dashed green line) since dividing by $G({\rm 217GHz})\simeq 0$
boosts the errors. Fig.~\ref{fig2}b demonstrates that the Comptonization parameters 
derived using weighted frequency averages (method A)
and the ILC map (method B) are fully compatible. To avoid
overcrowding the plot, error bars are not shown. The red line represents 
$\ln Y_{C,\nu, 500}=A+B\ln Y_{C,ILC,500}$ whose best fit parameters are
$A=0.001\pm0.007$ and $B=0.97\pm0.03$, consistent 
with the expected values of $A=0$ and $B=1$ at the $1\sigma$ 
confidence level (CL). 
In Fig.~\ref{fig2}c we compare the error using each estimator to show
that while the values of $Y_{C,500}$ measured by both methods are
comparable, the errors given by the ILC method (red histogram)
are slightly smaller. 
This is logical since the 217~GHz channel is used to construct the ILC map 
while it is not used in the weighted average of eq.~(\ref{eq:weighted_average}).
Therefore, in our subsequent analysis we will quote the results using the ILC method.

\section{Scaling Relations.}\label{sec4}

To determine the scaling relation between the Comptonization parameter measured 
from the ILC map as described above $Y_{C,SZ}$ and the value predicted using 
X-ray data $Y_{C,Xray}$ we will use the isothermal $\beta=2/3$ profile and 
the universal pressure profile with the two set of parameters given in 
Table~\ref{tab:gnfw_pars}. We will compare the measured and the predicted 
values at two angular scales, $\theta_X$ and $\theta_{500}$. We subdivide our 
sample in five bins to study if the scaling relations evolve with 
redshift. All bins have width $\Delta z=0.05$ except the last one where 
$\Delta z=0.1$ since only 19 clusters have $z\ge 0.25$.
The average cluster properties of the full sample and the different
subsample are given in Table~\ref{tab:cluster_properties}.

\subsection{Self-Similar Scaling Relations.}\label{sec:regression}

While the dynamical evolution of clusters is dominated by the collapse
of the Dark Matter (DM) component, their observational properties are determined
by the physical processes undergone by the baryon component. In the self-similar
model all cluster observables scale with the cluster mass; in particular, 
the mass of the gas $M_g$, the luminosity $L_X$, the gas temperature $T_X$
and the Comptonization parameter $Y_C$ scale as: $M_g\sim M$, $L_X\sim E(z)^{7/3}M^{4/3}$,
$T_X\sim E(z)^{2/3}M^{2/3}$ and $Y_C\sim E(z)^{2/3}M^{5/3}$ \citep{kaiser_86}.
In this expression, $E(z)$ is the Hubble function in units of the Hubble constant
today. In the Kaiser model, $Y_C\propto E(z)^{9/4}L_X^{5/4}$ \citep{Maughan2007}.
To facilitate the comparison with earlier results 
we will fit scaling relations of the form $E(z)^\gamma[D_A^2Y_{C,SZ}]=
10^A[E(z)^\kappa X/X_0]^B$. Specifically,
\begin{align}
\label{eq:yc_scaling}
& E(z)^{\gamma_Y}\frac{D_A^2 Y_{C,SZ}}{{\rm Mpc}^2}=10^A 
\left(\frac{D_A^2Y_{C,Xray}}{10^{-4}{\rm Mpc}^2}E(z)^{\kappa_Y} \right)^B,\\[0.3cm]
& E(z)^{\gamma_L}\frac{D_A^2 Y_{C,SZ}}{\rm Mpc^2}=10^A 
\left(\dfrac{L_X}{
7\times10^{44}\, {\rm erg/s}}E(z)^{\kappa_L}\right)^B ,
\label{eq:lx_scaling}
\end{align}
where the luminosity is measured in the $[0.1-2.4\, {\rm keV}]$ band.
The chosen normalizations are those of \cite{planck_early_XI}. In 
eq.~\eqref{eq:yc_scaling}, $\gamma_Y$ and $\kappa_Y$ parametrize any possible
redshift dependence due to observational biases on the measured cluster
X-ray and TSZ magnitudes.  
If there were no such systematics, $(\kappa_Y,\gamma_Y)=(0,0)$ and $(A,B)=(-4,1)$. 
In eq.~\eqref{eq:lx_scaling}, X-ray luminosity and Comptonization parameter
depend on electron density and temperature differently. In the self-similar model 
$(\kappa_L,\gamma_L)=(-7/3,-2/3)$ and $B=5/4$ while $A$ 
is determined observationally.  In this equation, the X-ray luminosity
normalization roughly corresponds to the mean X-ray luminosity of our sample. 
To simplify the analysis we will fix $(\kappa,\gamma)$ to their self-similar values.
To test the redshift evolution we will subdivide the sample in the redshift bins
described above and we will compute $(A,B)$ for each subsample. 

As mentioned in the introduction, the predictions of the self-similar
model do not coincide exactly with the observations. 
Deviations from the self-similar predictions are 
not unexpected. Since the concentration parameter that characterize the
DM profile depends on mass \citep{nfw1997}, if the gas is in hydrostatic
equilibrium within the DM potential well, one can expect that also the gas 
density and temperature profiles will deviate from the self-similarity assumed
in the Kaiser model. If we parametrize $M_g\sim M^{1+\alpha_g}$ and $T\sim M^{2/3+\alpha_T}$,
then the scaling relations become: $L\sim E(z)^{7/3}M^{4/3+2\alpha_g+\alpha_T/2}$,
$Y_C\sim E(z)^{2/3}M^{5/3+\alpha_g+\alpha_T}$ \citep{kravtsov_12}. 
This would correspond to $B=(5/3+\alpha_g+\alpha_T)/(4/3+2\alpha_g+\alpha_T/2)$
in eq.~\eqref{eq:lx_scaling}.  Then, a deviation from the value $B=5/4$ would
indicate to what extent the gas is better described by this extension 
of the self-similar model.

\begin{table*}
\begin{center}
\begin{tabular}{|c|c|c|c|c|c|c|}
\hline
& & & & & \\[0.8mm]
Subset & Ncl & $\bar{z}$ & $\bar{\theta}_{X}$ & $\bar{\theta}_{500}$ & 
$\bar{L}_X$ & $\bar{M}_{500}$ \\
& & & (arcmin) & (arcmin) & ($10^{44}$erg/s) & ($10^{14}M_\odot$)\\
\hline
\textbf{All Clusters}  & 560 & 0.11  & 6.93 & 12.02  &  2.45  & 3.20  \\[0.8mm]
\hline
$0.0<z<0.05$  & 95  & 0.035  & 10.15 & 21.60  &  0.59  &  1.40 \\[0.8mm]
$0.05<z<0.10$ & 217 & 0.074  &  7.30 & 12.60  &  1.37  &  2.45 \\[0.8mm]
$0.10<z<0.15$ & 107 & 0.123  &  6.02 & 8.81   &  2.19  &  3.30 \\[0.8mm]
$0.15<z<0.20$ & 79  & 0.169  &  5.06 & 7.65   &  4.14  &  4.80 \\[0.8mm]
$0.20<z<0.30$ & 62  & 0.24   &  4.67 & 6.46   &  7.38  &  6.42 \\[0.8mm]
\hline
\end{tabular}
\caption{Average properties of full cluster sample and subsamples.}
\label{tab:cluster_properties}
\end{center}
\end{table*}
\subsection{Fit methods.}\label{sec:regression_fit}

Linear fits are the most used regression algorithms; different statistical
estimators can be used to determine the intercept $A$, the slope $B$ and 
their respective uncertainties. If $(X_i, Y_i)$ are the data points
and $(\sigma_{X_i},\sigma_{Y_i})$ their respective uncertainties, 
the commonly used least squares method is a biased estimator. As
an alternative, \citep{akritas1996} introduced the Bivariate Correlated Errors
and intrinsic Scatter (BCES) that accounts for errors in both variables
and their intrinsic scatter with respect to the regression line. 
If there are low-precision measurements  that dominate 
over the other data points the method could be useless \citep{Tremaine2002}.
For comparison we use a Maximum Likelihood Estimator (MLE) 
introduced  by \citet{kelly2007}\footnote{Code downloaded 
from {\it http://idlastro.gsfc.nasa.gov/}} that also
accounts for correlated errors in both variables and their intrinsic scatter.
We computed the regression coefficients using both methods and found 
the results to be in excellent agreement. The MLE performs marginally 
better than BCES when the measurement errors and the intrinsic scatter 
are large.  Then, we will quote only the results estimated using the MLE technique.

When fitting a scaling relation, it is important to distinguish
between the {\it raw} scatter, $\sigma_{raw}$, the dispersion around the
best fit model, and the {\it intrinsic} scatter, $\sigma_{int}$, due to the
differences on physical properties of clusters.
The former is computed as the error weighted residual
\begin{equation}\label{eq:raw_scatter}
 \sigma_{raw}^2= \frac{1}{N_{cl}-2}\Sigma_{i=1}^{N_{cl}} \lambda_i(Y_i-BX_i-A)^2 ,
\end{equation}
where $N_{cl}$ is the number of clusters and the weights are given by 
\begin{equation}
 \lambda_i= \frac{N_{cl}\sigma_i^{-2}}{\Sigma_{i=1}^{N_{cl}} \sigma_i^{-2}},
\qquad \sigma_i^{2}=\sigma_{Y_i}^{2}+A\sigma_{X_i}^{2},
\end{equation}
while $\sigma_{int}^2=\sigma_{raw}^2-\sigma_{stat}^2$ is
the difference between the `raw' scatter and the statistical uncertainty 
($\sigma_{stat}^2=N_{cl}^{-1}\Sigma_{i=0}^{N_{cl}} \sigma_i^2$)
obtained by propagating the error on the measured quantities. A simple 
estimator of the intrinsic error is given by adding in quadrature
uniform values to the measured uncertainties of each individual cluster
and finding the value for which $\chi^2$ per degree of freedom is equal to unity, i.e., 
$\chi^2=\sum_i^{N_{cl}}(d_i-Y_i-BX_i-A)^2/(\sigma_{raw,i}^2+\sigma_{int}^2)\equiv N_{cl}-2$ 
(\citet{Maughan2007, Comis2011}). An alternative estimator has been used
in \cite{planck_early_X} that differs from the one described
previously on the fifth decimal place.

\subsection{Error bars.}\label{sec:errorbar}

For the $\beta$ model, the errors on $Y_{C,Xray}$ are dominated by 
the uncertainties in $T_X$ and $n_{e,0}$, while for the universal pressure
profile the dominant uncertainty is that of $r_{500}$ and, consequently,
of $M_{500}$. The error on $n_{e,0}$ is negligible compared with the 
error on $T_X$, a magnitude derived from a
scaling relation. In the case of the universal pressure profile,
we propagated the error on $M_{500}$ and added in quadrature
the uncertainty $\Delta_{par}$ due to the difference between the parameters 
of \citet{arnaud2010} and \citet{planck_int_05}. We 
estimated $Y_{C,Xray}$ using both sets of parameters, denoted by subindices
A and P, respectively, and we take this uncertainty to be the absolute value
of their difference: $\Delta_{par}=|Y_{C,Xray}^{P}-Y_{C,Xray}^{A}|$.
To estimate the errors on magnitudes derived from scaling relations
we generated 10,000 realizations of $T_X$ and $r_{500}$ for each cluster,
assuming that the errors on the parameters of their respective scaling 
relations were Gaussian distributed. On average, the relative error 
on $T_X$ was found to be $\sim 8\%$, while in 
$r_{500}$ was $ \sim 12\%$. The corresponding error on $M_{500}$ is 
$\sim 35\%$.  Since $\Delta_{par}$ contributes to the total relative 
error budget with less than $10\%$, the final error on the universal 
profiles was dominated by the uncertainty on the mass estimates.

\subsection{Effect of selection biases.}\label{sec:malmquist}

Selection biases affect X-ray flux limited samples in two ways:
Malmquist bias, due to higher luminosity clusters being preferentially selected 
out to higher redshifts and Eddington bias, due objects above the flux limit 
having above average luminosities for their mass as a result of the
intrinsic or statistical scatter in their luminosity for any given mass.
Scaling relations need to be corrected from these effects 
\citep{ikebe_02,stanek_06,pacaud_07,pratt2009,Vikhlinin2009,mantz_10,mittal_11}.
To determine the selection biases we follow the method outlined in 
\citet{bolocam_15}. We generated a sample of $5\times 10^4$ halo objects out to
$z\le 0.3$ by sampling the number density of halos of a given mass given
by \citet{tinker_08}.  All relevant cosmological parameters were fixed to 
the {\it Planck} measured values \citep{planck13_XVI}. To each halo mass we 
assign two magnitudes $X=[Y_{C,X-ray},L_X]$ using scaling relations; 
the Comptonization parameter comes from the relation
$Y_{C,X-ray,500}-M_{500}$ derived by \citet{arnaud2010} and
the X-ray luminosity from the $L_X-M_{500}$ relation given in \cite{lovisari_15}.
Both these relations are corrected from statistical biases. The
properties of individual clusters differ from one another
according to the measured intrinsic dispersion.
Then, we imposed the same flux cut than in the data and
selected a sample of 560 clusters with the
same redshift distribution than in the actual catalog.
We repeat the process till three hundred samples have been
selected.

The functional form of the scalings given in eqs.~(\ref{eq:yc_scaling}, 
\ref{eq:lx_scaling}) is $\log(Y_{C,SZ}/Y_{C,SZ,0})= B\dot\log(X/X_0) + A$.  
 We assign a value of $\log(X)$ from the scaling relations taking into
account the intrinsic scatter. The coefficients $(A,B)$ are
varied within predefined intervals. For the scaling relation of
eq.~(\ref{eq:yc_scaling}), the normalization and slope were varied 
in the range $A=[-4.50, -3.50]$ and $B=[0.85, 1.45]$ while for the 
scaling of eq.~(\ref{eq:lx_scaling}) the range of variation was
$A=[-4.50, -3.50]$ and $B=[1.00, 1.40]$, respectively. 
We fit the scaling relation to the 300 subsamples of simulated clusters for each
pair of grid points $(A,B)$ and we vary these coefficients in steps of 0.01. 
We use the same scaling relations to correct the selection biases determined
at the $\theta_X$ aperture. This aperture is slightly larger than
$\theta_{2500}$. The scalings measured at the latter aperture
are not very different from those at $\theta_{500}$ \citep{Bonamente2008}
and are well within the range defined above. In each realization, coefficients are
measured using the MLE estimator. Like in \cite{planck_early_XI}
slopes and amplitudes are adjusted until the mock observed samples match 
those recovered from the actual data. Then, the unbiased scaling relation
is that of the parent population. 

\section{Results and Discussion.}

We have used a sample of 560 X-ray selected clusters and the 2013 foreground 
cleaned Planck Nominal maps to determine two scaling relations,
$Y_{C,SZ}-Y_{C,Xray}$ and $Y_{C,SZ}-L_X$, using the BCES(Y$|$X) and the
MLE regression methods but the differences are below 1\%
and only the results from the latter method will be quoted. 
The uncertainties were determined by 10,000 
bootstrap re-samplings \citep{akritas1996, kelly2007}. 

\begin{table*}
\begin{center}
\caption{Scaling relations with MLE regression coefficients (eqs.~\ref{eq:yc_scaling},
\ref{eq:lx_scaling}) corrected and not corrected from statistical biases.}
\label{tab:fit_results}
\begin{tabular}{lcccc|cccc}
\hline
\multicolumn{1}{l}{Relation } & \multicolumn{8}{c}{Angular Size} \\[0.1cm]
\multicolumn{1}{c}{ } & \multicolumn{4}{c|}{$\theta_X$} & \multicolumn{4}{c}{$\theta_{500}$}\\
\cline{2-9} \\[-0.1cm]
\multicolumn{1}{c }{ } & \multicolumn{1}{c}{$A$} & \multicolumn{1}{c}{$B$} &
\multicolumn{1}{c|}{$\sigma_{\rm raw}$ } & \multicolumn{1}{c|}{$\sigma_{\rm log_i.}$ } &
\multicolumn{1}{c}{$A$} & \multicolumn{1}{c}{$B$} &
\multicolumn{1}{c|}{$\sigma_{\rm raw}$ } &
\multicolumn{1}{c|}{$\sigma_{\rm log_i}$ }\\
\hline
$^\blacktriangle Y_{C,SZ}$--$Y_{C,Xray}$ &  
$-4.09\pm0.04$ & $1.07\pm0.03$ & 0.393 & $0.392$
&$-4.22\pm0.04$ & $0.95\pm0.04$ & 0.362& $0.360$\\[0.5cm]
$^\bigstar Y_{C,SZ}$--$Y_{C,Xray}$ & 
$-4.00\pm0.03$ & $1.22\pm0.03$ & 0.467 & $0.465$
&  $-4.09\pm0.03$ & $1.16\pm0.05$ & 0.478 & $0.476$\\ [0.5cm]
$^\blacksquare Y_{C,SZ}$--$Y_{C,Xray}$ & 
$-3.95\pm0.03$ & $1.21\pm0.03$ & 0.474& $0.472$
&  $-4.04\pm0.03$ & $1.16\pm0.05$ & 0.478 & $0.476$\\ [0.5cm]
$^\Diamond Y_{C,SZ}$--$L_X$[0.1-2.4]keV & 
$-4.27\pm0.04$ &  $1.19\pm0.05$ & $0.493$ & $0.491$
& - &  -  & -  & -\\
\hline
\hline
{\bf Selection Bias Corrected Relations} &   & & & & & &   &\\
\hline
$^\blacktriangle Y_{C,SZ}$--$Y_{C,Xray}$ &  
$-3.99\pm0.04$ & $1.11\pm0.04$ & 0.393 & $0.392$
&$-4.10\pm0.04$ & $0.97\pm0.04$ & 0.362& $0.360$\\[0.5cm]
$^\bigstar Y_{C,SZ}$--$Y_{C,Xray}$ & 
$-3.85\pm0.03$ & $1.27\pm0.03$ & 0.467 & $0.465$
&  $-3.96\pm0.03$ & $1.20\pm0.05$ & 0.478 & $0.476$\\ [0.5cm]
$^\blacksquare Y_{C,SZ}$--$Y_{C,Xray}$ & 
$-3.81\pm0.03$ & $1.25\pm0.03$ & 0.474& $0.472$
&  $-3.91\pm0.03$ & $1.21\pm0.05$ & 0.478 & $0.476$\\ [0.5cm]
$^\Diamond Y_{C,SZ}$--$L_X$[0.1-2.4]keV & 
$-3.98\pm0.05$ &  $1.25\pm0.04$ & $0.493$ & $0.491$
& - &  -  & -  & -\\
\hline
\end{tabular}
\end{center}

\begin{flushleft}
$^\blacktriangle$: X-ray pressure profile from the $\beta=2/3$-model.\\
$^\bigstar$: X-ray pressure profile from \citet{arnaud2010}.\\
$^\blacksquare $: X-ray pressure profile from \citet{planck_early_XI}.\\
$^\Diamond$: We assumed an error of 10\% on the X-ray luminosity.\\
\end{flushleft}
\end{table*}

\subsection{The SZ-Xray scalings from pressure profiles.}

\begin{figure*}
\epsfxsize=.9\textwidth \epsfbox{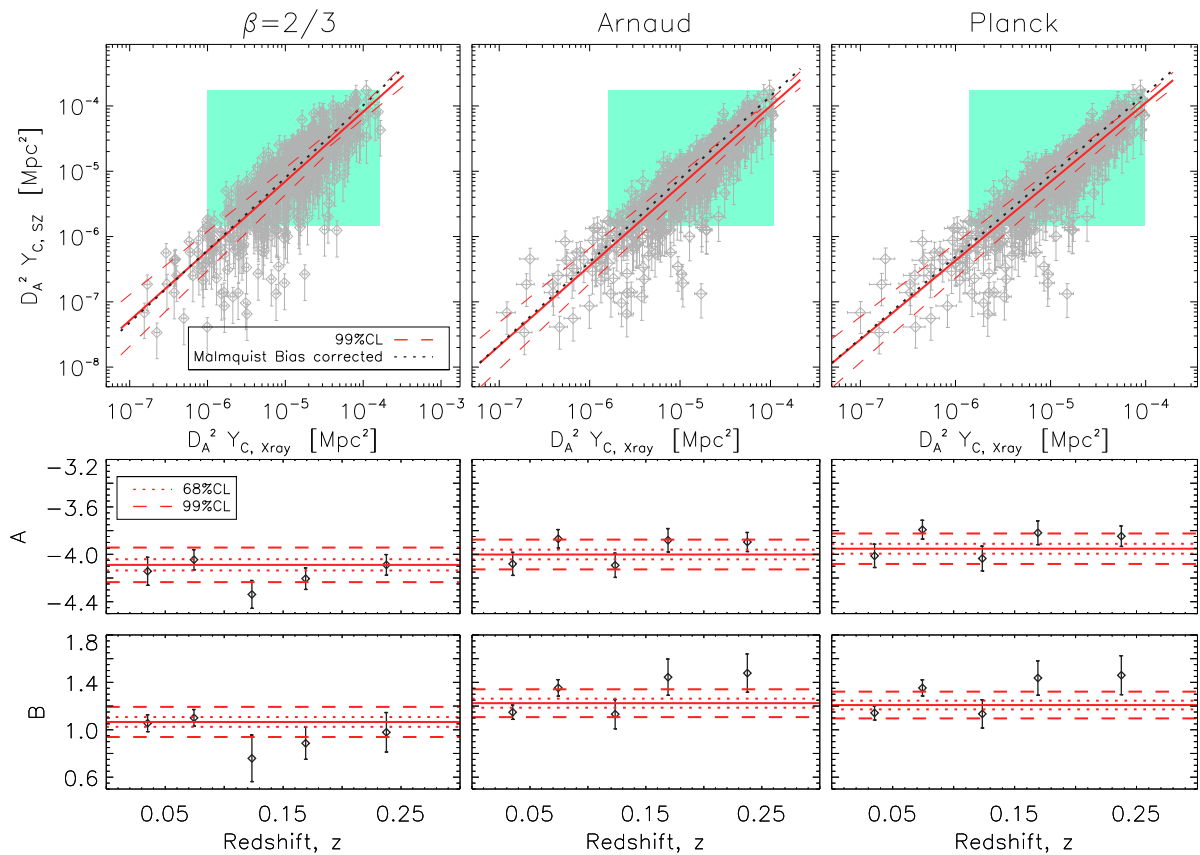}\\
\caption{Top panels: 
Scaling relation between the Comptonization parameter $Y_{C,Xray}$ 
predicted using X-ray data, the $\beta=2/3$-profile and the
universal profiles with \citet{arnaud2010} and 
\citet{planck_int_05} parameters and the same magnitude 
$Y_{C,SZ}$ measured in the ILC map. The signal was estimated/measured
on discs of radius $\theta_X$. The solid red line represents the 
best fit, the dashed lines the uncertainty at the 68\% CL on the fit and
the dotted line the fit corrected from the statistical biases as indicated
in Sec.~\ref{sec:malmquist}.
The cyan square indicates the region occupied by the clusters used by the
{\it Planck} Collaboration. Middle and bottom panels:
intercepts and slopes of the scaling relation on the 
cluster redshift sub-samples. The solid, dotted and dashed red lines
indicate the best fit, 68\% and 99\% CL of the best model
represented in the top panels.} 
\label{fig3}
\end{figure*}

\begin{figure*}
\epsfxsize=.9\textwidth \epsfbox{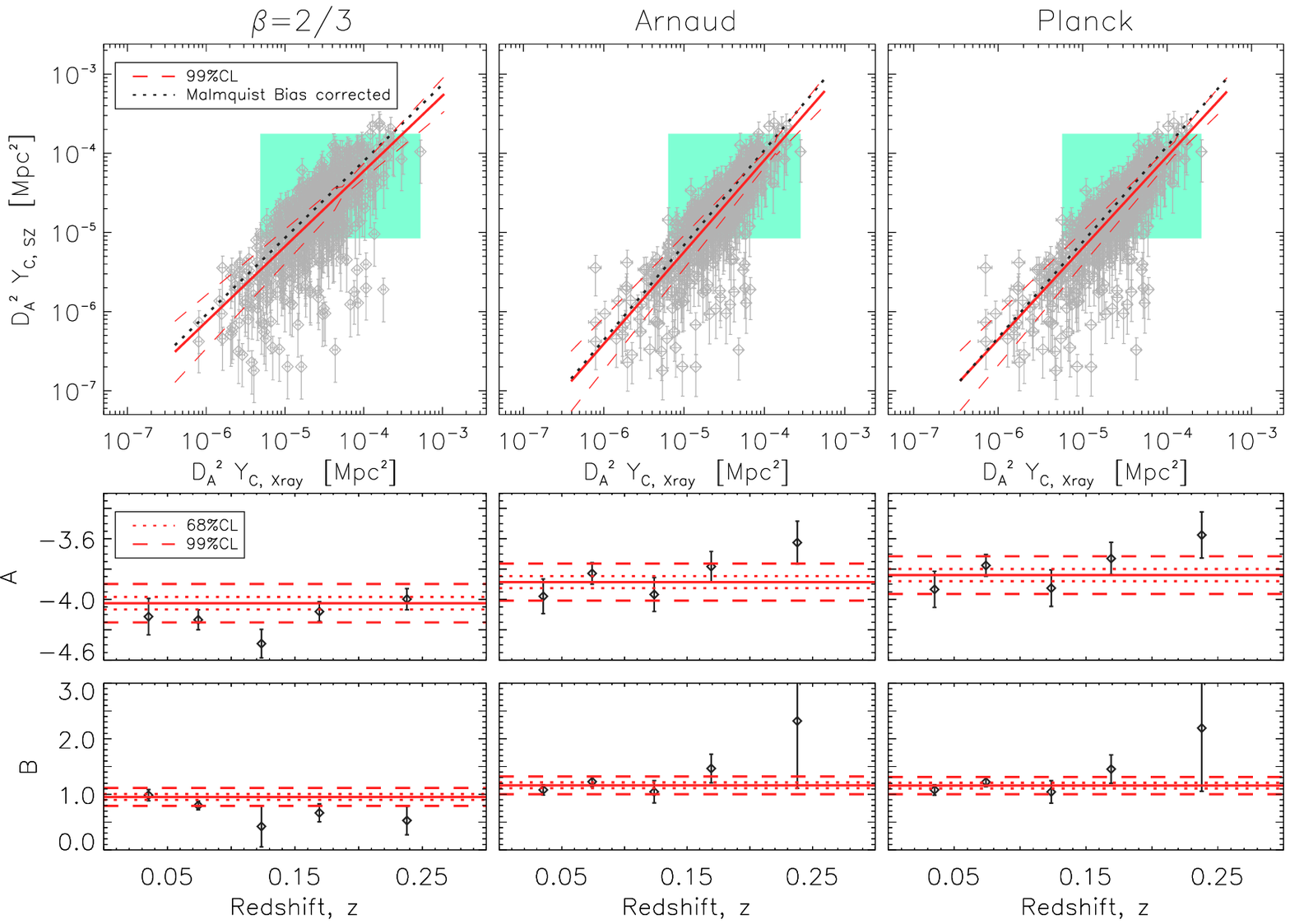}\\
\caption{Same as in Fig.~\ref{fig3} except that the magnitudes are averaged 
on discs of radius $\theta_{500}$.}
\label{fig4}
\end{figure*}

In Figs.~\ref{fig3} and \ref{fig4} we represent the data and scaling parameters 
of the $Y_{C,SZ}-Y_{C,Xray}$ relation. The Comptonization parameters were 
obtained by averaging the temperature anisotropies on discs on angular 
size $\theta_X$ (Fig.~\ref{fig3}) and $\theta_{500}$ (Fig.~\ref{fig4}).
$Y_{C,SZ}$ was measured directly on the foreground cleaned {\it Planck} 
Nominal maps while $Y_{C,Xray}$ was computed
using the X-ray profiles described in Sec.~2: $\beta=2/3$ model and
universal profile with \cite{arnaud2010} and \cite{planck_int_05}
parameters. In the top three panels we plot the full data and 
its error bars. The solid red line corresponds to the best fit
and the dashed lines to scaling relation with the parameters modified by
$1\sigma$. The dotted line represents the scaling relation 
corrected of the statistical biases as described in Sec.~\ref{sec:malmquist}.
The cyan square shows the region of the parameter space occupied by the clusters 
used in \cite{planck_early_XI}. 

In Table~\ref{tab:fit_results} we give the best fit parameters and
the raw and intrinsic errors for the full catalog. The intrinsic scatter is 
always smaller but similar to the raw scatter, demonstrating that
the uncertainties on the scaling relations are due to the physical differences 
within the cluster population and not to the statistical uncertainties on
the measured magnitudes. For comparison
we give the parameters of the scaling relation
derived from the data and corrected from the statistical biases as
indicated in Sec~\ref{sec:malmquist}. The raw and intrinsic errors are
identical (differences are seen only at the fourth
decimal place) since, by construction, the mock catalogs of simulated clusters
used to correct for selection biases had the same dispersion than the data.
 At the $\theta_X$ aperture, the bias corrected relation shows that the 
intercept and slope of the $\beta=2/3$ model are $A=-3.99\pm 0.04$
and $B=1.11\pm 0.04$, deviating by about $2.7\sigma$ from the 
expected values of $A=-4$ and $B=1$. The deviations are larger when 
the bias corrected relations of the 
universal profile are used. Evaluating magnitudes at the $\theta_{500}$
aperture, the intercept from the $\beta=2/3$ model deviates by 
a $2.5\sigma$ but the slope is compatible with $B=-1$ while the universal
profile predict slopes that deviate  by more than $4\sigma$ but the intercepts
are closer to the expected value. To quantify these deviations, we compute 
the mean Comptonization parameter weighted by the angular extent of each cluster,
$\bar{Y}_C=\sum_i(Y_{C,i}\theta_i^2)/\sum_i\theta_i^2$.
At $\theta_X$ and $\theta_{500}$, this average is
$\bar{Y}_{C,Xray}=(1.04,1.17)\bar{Y}_{C,SZ}$ for the $\beta$ model
and is $\bar{Y}_{C,Xray}=(1.15,1.21)\bar{Y}_{C,SZ}$ for the universal
profile, respectively.
Then, on average the $\beta=2/3$ model correctly predicts the TSZ amplitude
at $\theta_X$ (4\% excess) but overpredicts it beyond this radius (17\% excess)
as already demonstrated by \citet{atrio2008}. The universal
profile overpredicts the signal by a 21\% at $\theta_{500}$, contrary to
earlier findings by the Planck Collaboration \citep{planck_early_XI}.
Similar results were obtained using a universal profile
with the \citet{planck_int_05} parameters.

We can verify if the discrepancy between the prediction of the universal 
pressure profile and the measured anisotropy is related to the cooling process 
of the ICM by dividing clusters into {\it cool-core} and 
{\it non cool-core} systems.  We adopt the \cite{oHara2006} classification: 
CC clusters are those with central cooling times below the Hubble time ($t_H$).
We used the central electron densities, X-ray temperatures listed in our 
catalog and the definition of cooling time $t_{cool}=8.5\times10^{10}\rm{yr}
\left(n_e/10^{-3}\rm{cm}^{-3}\right)^{-1}\left(T_X/10^8\rm{K}\right)^{1/2}$
by \citet{Sarazin1988} to distinguish between CC and NCC clusters.
The error on this cooling time was
estimated by propagating the uncertainty of the X-ray temperature.
In total, only 63 clusters in our sample were cool-core clusters.
Reproducing the analysis, we found no significant difference
with the results obtained using the full sample. 
When analyzing both subsamples separately, the results were similar in each
subset; the only noticeable effect was the expected increment on the statistical
uncertainty due to the smaller number of clusters, but the intrinsic scatter 
was still the largest component of the total scatter. Therefore, we can not
ascribe the discrepancy to the presence of CC systems in our catalog.

At $\theta_{500}$ our result using 
the universal profile is $(A,B)=(-3.96\pm 0.03, 1.20\pm 0.05)$, rather different
from $(A,B)=(-3.91\pm 0.01, 0.96\pm 0.04)$ found by \cite{planck_early_XI}.
To understand the source of this discrepancy, we repeat the analysis using only 
the 358 in the same mass and temperature range than the ones used by the Planck 
Collaboration, $M_{500}=[2 - 20]\cdot 10^{14}\rm{M}_\odot$ and $T_X=[3-14]$ keV. 
 For this subsample we obtain $(A,B)=(-3.99\pm 0.03, 1.08\pm 0.05)$
and the discrepancy is reduced to less than $3\sigma$. 
When averaging over the clusters
angular extent, we obtain  $\bar{Y}_{C,Xray}=1.035\bar{Y}_{C,SZ}$
at $\theta_{500}$,  an excess of less than 4\%.
The intrinsic scatter in this subsample is also reduced 
from $\sigma_{log_{int}}=0.476$ to $\sigma_{log_{int}}=0.17$, i.e.
the dispersion around the best fit is much smaller, but it is still greater 
than the one in the {\it Planck} analysis, $\sigma_{log_{int}}=0.10$. The small
difference with respect to the Planck result
suggests that the universal pressure profile with a 
unique set of parameters describes well clusters with masses 
$\gsim 10^{14}M_\odot$ and X-ray temperatures $\gsim 3$keV,
comparable to the range analyzed by \citet{arnaud2010}) while
for less massive clusters the profile is not as accurate. 
This confirms the trend found by \citet{lovisari_15} with a much
smaller sample: the more massive systems had a shallower slope
and could be an indication that the $L_X-M$ relation is gradually 
steepening when moving toward the low-mass objects suggesting that
a simple power law cannot be used to convert the measured luminosities 
into masses.

The diamonds and error bars in the 
middle and bottom panels of Figs.~\ref{fig3},~\ref{fig4} show the
value of intercept and slope $(A,B)$ in the five redshift bins described in 
Sec.~\ref{sec4} (not corrected from statistical biases). The solid
dotted and dashed line show the mean, 68\% and 99\% CL of the full
sample.  There is no clear trend in either parameter for the three 
pressure profiles considered. The value of the coefficients A and B
at each redshift is always compatible with the fit of the full sample,
indicating that the scaling does not evolve with time and confirming the results
of \cite{planck_early_XI}.  Nevertheless, 
we can not disregard the importance of statistical biasing effects
since in the high redshift bins massive clusters are better represented compared 
with the low mass ones. Unfortunately, our redshift bins contain few clusters and 
the error bars on the measured values are consequently large so no statistical 
analysis could be made.

\subsection{The $Y_{C,SZ}$-Xray luminosity relation.}
\begin{figure}
\epsfxsize=.45\textwidth \epsfbox{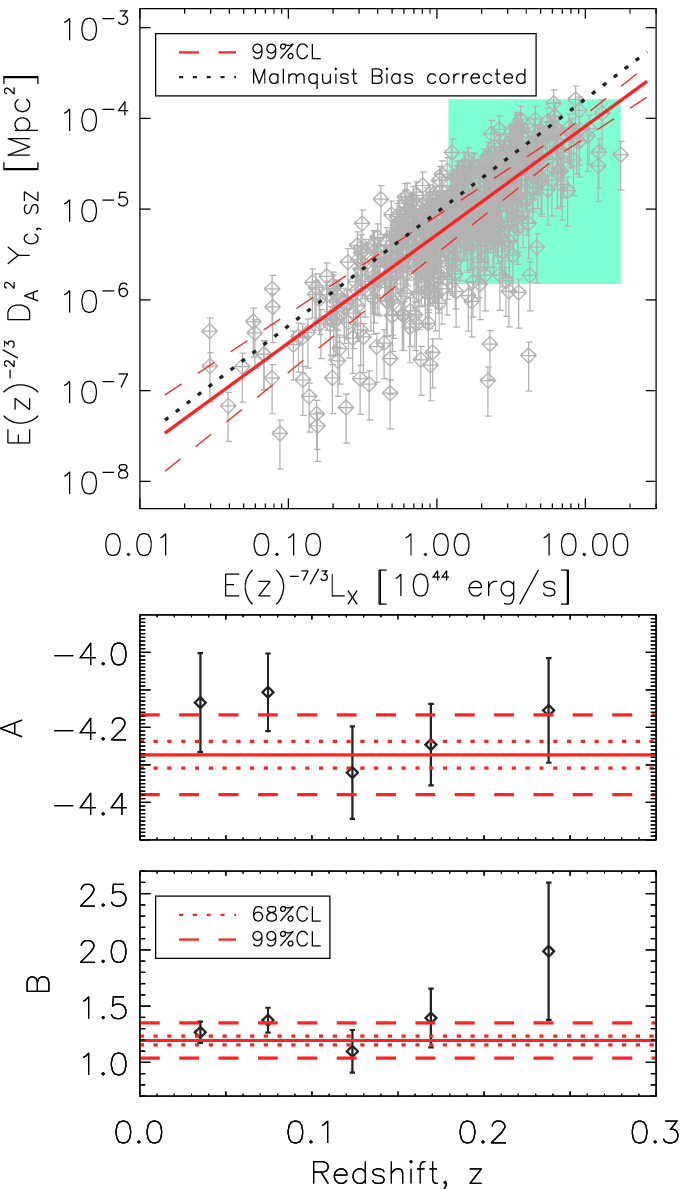}
\caption{The $Y_{C, SZ}-L_X$ scaling relation at $\theta_X$.
Panels and lines follow the same conventions as in Fig.~\ref{fig3}.}
\label{fig5}
\end{figure}

Equally important is the scaling relation of the Comptonization parameter with
X-ray luminosity since we can test the extensions of the self-similar model
described in Sec.~\ref{sec:regression}. In this case, we limit our
study to $\theta_X$, the aperture at which the X-ray magnitudes
have been measured. In Fig.~\ref{fig5} we present the data, the 
best fit (solid line), the best corrected from statistical biases
(dotted line) and the $1\sigma$ deviations from the best fit.
We quote results assuming a 10\% uncertainty in 
the X-ray luminosity of all clusters, but the results remain unchanged 
if we conservatively increase the relative errors to 20\%.
The slope and intercept are given in Table~\ref{tab:fit_results}.  

The measured correlation parameter  corrected from
statistical biases is $B=1.25\pm 0.04$, fully compatible 
with the self-similar value $B=5/4$. 
The deviations from self-similarity
are compatible with zero: 
if we assume $\alpha_T=0$ then $\alpha_g\simeq 0.0\pm 0.03$
while if $\alpha_g=0$ then $\alpha_T\simeq 0.0\pm 0.15$.
Our value is not directly comparable
to $\alpha_g=0.13\pm 0.03$ found by \cite{lin_12} using 
94 clusters in the range $z=0-0.6$ since magnitudes were evaluated
at $\theta_{500}$.
In \cite{planck_early_X} a similar analysis was carried out 
using a catalog of $\sim 1600$ clusters up to redshift $z\sim 1$. 
For this sample, the measured value was $B\simeq 1.095\pm 0.025$.
Again, the value is not directly comparable
to ours since X-ray luminosities are measured at $\theta_{500}$ and 
the data was binned in luminosity to reduce errors. 

In Fig.~\ref{fig5} also we present the parameters measured at different
redshift using the same notation as in Figs.~\ref{fig3} and \ref{fig4}. Like
before, we found no indication of redshift evolution, confirming the
results of \citet{Melin2011} and \citet{planck_early_XI}.
However, we have to consider that our redshift range could 
be too small to be sensitive to any hypothetical evolution.
For instance, \citet{Maughan2007} used a sample of 115 clusters at redshifts
$z=0.1-1.3$ to carry out his analysis. Computing the
cluster mass from a $M-T_X$ relation and using a universal
gas to dark matter fraction, he estimated the Comptonization
parameter for each cluster and determined 
the $Y_C-L_X$ relation for different apertures. Since
$Y_C$ was not measured from CMB temperature anisotropy
maps, his results are not directly comparable with ours.
Nevertheless, he systematically finds the exponent to be $B\le 1.1$.
Since his cluster sample extends to a greater redshift range than
the one used here, it would be interesting to test if 
his results, when compared to ours, are an indication of
the time evolution of the scaling parameters.

\section{Conclusions.}

We have fitted scaling relations between the Comptonization parameter
predicted using X-ray data and the measured X-ray luminosity with 
the one measured from foreground cleaned {\it Planck} 
2013 Nominal maps using a sample of 560 X-ray selected clusters. 
Prediction and measurement are compared at two angular scales,
$\theta_X$ that correspond to the region that emits 99\% of the X-ray flux
and $\theta_{500}$, the scale at which the cluster density is 500 times
the critical density. Our catalog contains cluster and rich groups
with masses $M_{500}\gsim 10^{13}M_\odot$, one order of magnitude below the mass 
range explored by the Planck Collaboration. We found that the Comptonization 
parameter $Y_{C,Xray}$ predicted using the $\beta=2/3$ model agrees with 
the measured value within $\theta_X$ but overestimates it at $\theta_{500}$ 
indicating that clusters are not isothermal. We also show that the
Universal profile with either \citet{arnaud2010} or \citet{planck_int_05} parameters
overestimates the SZ anisotropy.  Averaged over our cluster sample, we
find an excess ranging from  15\% to 21\% depending on the aperture and
the set of parameters used. This is slightly lower than the 30\% 
required to explain the TSZ power determined by WMAP \citep{komatsu2011}. 
We verified that the discrepancy is not due to the presence of
CC clusters. The discrepancy is greatly reduced if 
the analysis is restricted to clusters with $M_{500}\gsim 10^{14}M_\odot$,
suggesting that the low and high mass systems are not well described by a
universal pressure profile with the same set of parameters, an indication
that the dynamical evolution of baryons in low and high mass systems
was different. This conclusion supports the \cite{lovisari_15} suggestion 
of a brake in the $L_X-M$ relation after correcting for selection
biases. Then a simple power law can not be used to describe both clusters
and groups and to translate X-ray luminosities into masses.
Finally, we have also shown that the relation of the Comptonization parameter
averaged over the region that emits 99\% of the X-ray flux
and the X-ray luminosity are consistent with the prediction of the self-similar
model.

We found that the scaling relation $Y_{C,SZ}-L_X$ is
consistent with the self-similar model described in Sec.~\ref{sec:regression}.
If the temperature scales with mass as in the self-similar model ($\alpha_T=0$)
then deviations of the scaling of the gas mass with the total mass are
compatible with zero ($\alpha_g=0.0\pm 0.03$) and is in tension with
the value of $\alpha_g=0.13\pm 0.03$ measured by \cite{lin_12}, although
the two results are not directly comparable since $\alpha_g$ was determined
from scaling relations measured at different apertures.

We tested the redshift evolution 
by dividing the catalog in five redshift subsamples. We found no evidence
of evolution within $z<0.3$ in the two scaling relations analyzed.
A comparison with the earlier results of \cite{Maughan2007} is not
straightforward since this author did not obtain $Y_{C,SZ}$ from
CMB data but used scaling relations. His sample of 115 clusters 
includes systems with $z=0.1-1.3$ and he finds the exponent to be $B\le 1.1$, 
depending on the angular scale used. The difference is significant enough to  
suggest that the scaling relations could be evolving in time but our
sample is not deep enough to be sensitive to the effect. This question
will require a separate study.

\section*{Acknowledgments}
We warmly thank H. Ebeling and D. Kocevski for sharing with us their
cluster X-ray data and the referee for helping us to improve the paper. 
IDM acknowledges financial 
support from the University of the Basque Country UPV/EHU under the program
"Convocatoria de contrataci\'{o}n para la especializaci\'{o}n de personal 
investigador doctor en la UPV/EHU 2015", and from the Spanish Ministry of 
Economy and Competitiveness through research project FIS2014-57956-P 
(comprising FEDER funds);  FAB acknowledges financial support from the 
Spanish Ministerio de Educaci\'on y Ciencia (grant FIS2015-65140-P)
and to the ``Programa de Profesores Visitantes Severo Ochoa'' of 
the Instituto de Astrof\1sica de Canarias.

\label{lastpage}
\end{document}